# Title: Investigating the dissemination of STEM content on social media with computational tools

**Authors:** Oluwamayokun Oshinowo[1,2,3]†, Priscila Delgado[1,2,3]†, Meredith Fay[1,2,3]†, C. Alessandra Luna[1,2,3], Anjana Dissanayaka[1,2,3], Rebecca Jeltuhin[1,2,3], David R. Myers[1,2,3]*

†These authors contributed equally to this work

*Corresponding author. Email: david.myers@emory.edu

**Affiliations:**
[1]The Wallace H. Coulter Department of Biomedical Engineering, Georgia Institute of Technology and Emory University, Atlanta, Georgia
[2]Department of Pediatrics, Aflac Cancer Center and Blood Disorders Service, Children's Healthcare of Atlanta, Emory University School of Medicine, Emory University, Atlanta, Georgia
[3]Parker H. Petit Institute of Bioengineering and Bioscience, Georgia Institute of Technology, Atlanta, Georgia

**Contact information:**
Oluwamayokun Oshinowo - ooshin2@emory.edu
Priscilla Delgado - priscilla.delgado@emory.edu
Meredith Fay- meredith.fay@gapjunctiondata.com
C. Alessandra Luna- luna.cecilia@emory.edu
Anjana Dissanayaka- anjana.dissanayaka@emory.edu
Rebecca Jeltuhin- rebecca.jeltuhin@emory.edu
David R. Myers- david.myers@emory.edu

# Title: Investigating the dissemination of STEM content on social media with computational tools

**Abstract:** Social media platforms can quickly disseminate STEM content to diverse audiences, but their operation can be mysterious. We used open-source machine learning methods such as clustering, regression, and sentiment analysis to analyze over 1000 videos and metrics thereof from 6 social media STEM creators. Our data provide insights into how audiences generate interest signals(likes, bookmarks, comments, shares), on the correlation of various signals with views, and suggest that content from newer creators is disseminated differently. We also share insights on how to optimize dissemination by analyzing data available exclusively to content creators as well as via sentiment analysis of comments.

**Introduction:**

Social media platforms such as Instagram, TikTok, and YouTube provide a new venue to promote STEM education, inspire the next generation of diverse scientists, and share knowledge to lower barriers to academia(*1-3*). Unlike many existing venues, social media is broadly accessible and not limited to those with significant resources devoted to their education. Content can be quickly disseminated to large diverse audiences of all ages and backgrounds(*4*). For example, disseminating content through TikTok, a platform with 1.6 billion active users (~68% below the age of 34), a video can potentially reach tens of millions of video views within 1 to 7 days. Contextualizing this number of views from TikTok, the most viewed article in JAMA in 2022 was viewed 941,313 times(*5*), with majority of other articles across other popular journals on average having views that are orders of magnitude lower.

Broadly disseminating STEM content to the broader public can be challenging for many new and established scientific content creators. While a single video has the potential to reach tens of millions of viewers, many videos only receive hundreds or thousands of views. We know that social media platforms use algorithms to curate content and/or provide recommendations to a user based on how they interact with a video. Some interactions are public, for example, on TikTok, each video displays the number of likes, comments, bookmarks, and shares. Other user interactions, such as the type of content watched, or the video viewing time remain private and only measured by the platform with some metrics available to the content creators. We also know that some platforms also make recommendations based on content that was of interest by other viewers that tend to have similar content preferences(*6*). However, these algorithms are all proprietary, prompting much speculation on what drives dissemination. Anecdotes on a variety of topics have been shared, from the "best" time to release new content to the relative importance of the various interest signals.

**Methods:**

Excellent peer-reviewed algorithms available to all researchers via open source, freely available Python programming language (https://www.python.org/, version 3.10) libraries provide mathematical context for the TikTok engagement data presented here. Jupyter notebook of methods and codes for our specific computational analysis are available at https://github.com/SL2lab/Oshinowo_et_al_2024

Correlation analysis:

Correlation matrices implemented using Pandas (https://pandas.pydata.org/, version 2.2.0)(*7*) are a statistical technique used to evaluate the relationship between two variables in a data set. In the produced table, every cell contains a Pearson correlation coefficient(*8*). +1 is considered a strong positive association between pairwise variables, 0 a neutral relationship, and -1 a strong inverse relationship.

Clustering and regression analysis:

Clustering analyses are a label-free machine learning method designed to find natural groupings with data sets. Here, k-means algorithms are implemented using scikit-learn (https://scikit-learn.org/stable/, version 1.4)(*9*). K-means algorithms(*10*), understood to be a robust general approach to clustering, group data points into a specified k number of clusters in which each observation belongs to the cluster with the nearest mean. Regression analyses, also implemented with scikit-learn, are used to mathematically characterize the value of a dependent variable (y-axis) based upon the value of an independent variable (x-axis). Here, we performed paired value analysis with views as the dependent variable.

Sentiment analysis of words and comments:

Sentiment analysis(*11*) as implemented using the Natural Learning Toolkit (https://www.nltk.org/, version 1.8)(*12*), a subset of the field of natural language processing, classifies individual words or groups of words (here, a comment) as having a positive (value greater than zero with a maximum value of 1), neutral (value of zero), or negative (value less than zero with a maximum value of -1) polarity. Word clouds as created by library WordCloud (https://github.com/amueller/word_cloud, version 1.8), groupings of words with size of text indicating relative frequency, are used to further quantify and visualize raw comment data.

## Results:

### Which interest signals matter?

As scientific content creators, we sought to use an evidence-based approach to gain insights into how user signals influence the dissemination (views) and impact of new content, specifically by using machine learning, the use of statistical algorithms that can learn and adapt without explicit instruction. Our approach can be used across platforms, although we focus here on TikTok as it has a rich set of publicly available interest signals (including likes, comments, shares, bookmarks, and views) available for each video. In addition to analyzing our own TikTok channel, the Scieneers (Figure 1a), we also analyzed the collective works of five additional content creators with varying popularity as measured by their followers, who are social media users that elect to prioritize content from that creator. This diverse set of content creators work across a variety of scientific disciplines in higher education, and each began their STEM outreach efforts using short-form content (Figure 1a). We did not include the excellent work of public-facing educators such as Bill Nye, Neil deGrasse Tyson, Michio Kaku, and Hank Green, since their celebrity status could skew our analysis.

We began our analysis by collating the public interest signal data for 1141 videos (excluding branded or sponsored content, as the algorithm treats these videos differently) into a single data set. We then calculated pairwise Pearson correlation coefficients(*7*)'(*8*), a measure of the strength and direction of the linear relationship between two continuous variables, for all interest signals, including views. These videos had each been on TikTok for more than 7 days, a timeframe where we have found the number of views and interest signals tends to stabilize for the vast majority of content. Our analysis allowed us to identify and visually represent any correlation between likes, comments, bookmarks, and shares with total views (Figure 1b). We found that all interest signals had a moderate or strong correlation to the number of views. The strongest correlation with views occurred with likes (0.95), demonstrating that the number of likes is the most predictive of views, although this data suggests that all interest signals are important. We also found that there was a strong correlation between likes, bookmarks, and comments suggesting that users will use multiple interest signals (Figure 1b). These trends were generally consistent between all creators with some more subtle differences. For example, with our account (Scieneers), comments had a substantially lower correlation with views (.61 versus the pooled .81) (Supplementary Figure 1).

**Are all videos treated equally?**

After establishing the importance of all interest signals, we investigated if the social media algorithm parameters driving views were applied to all videos equally. To that end, we used clustering(*9, 10*), an unsupervised machine learning technique that can pool single data points (videos) described by multiple features (likes, comments, shares, and bookmarks) and partition each video into mathematically optimized groupings (clusters) based on the feature similarity (Figure 1c, Supplementary Figure 2). Based on mathematical determination of an optimal number of clusters, we identified four groupings of videos and performed linear regressions. All creators had at least 1 video in cluster 2 (>1 million views), although this was a minority of their content. Clusters 2-4, accounted for 7.4% of all analyzed videos, and were composed of videos with very high view counts (>1 million). Interestingly, clusters 3 and 4, which correspond to the most viewed videos (>3 million views), were populated only by the two creators with large follower counts (>500k). We speculate that the content creator's follower count may play a significant role in driving content dissemination as clusters 2-4 were mostly populated from accounts with a higher number of followers. Additionally, similar to the correlation analysis, we found with linear regression that likes had the strongest correlation with views and although not used for the regression analysis, it is clear that branded content is treated differently by the algorithm (Figure 1c, Supplementary Figure 2).

For our team and other newer creators, most videos are grouped into cluster 1 (92.6% of videos). Interestingly, the strength of the correlations between interest signals and views was different for individual clusters than for all videos considered together. In cluster 1, only likes had a strong correlation ($r^2$=0.86) with views as measured by linear regression (Figure 1d), a statistical method used to model the relationship between a dependent variable and one or more independent variables by fitting a linear equation to the observed data. Effectively, this helps new creators because they can best understand the impact of their work by looking at a single interest signal. Our analysis also provided information on the relative use of the various interest signals by audiences. Bookmarks and shares are used ~10x less than likes, while comments are comparably rare and used ~100x less than likes. Noting our earlier finding that interest signals tend to be highly correlated to one other, this means that a typical user will like videos that they find interesting, and then bookmark, comment, or share a subset of those videos. We find that shares are the least correlated interest signal. We speculate that users may opt to share videos that they themselves do not find entertaining, but that a friend may find interesting. Noting that there is a significant amount of variability in the interest signals and views, our data also shows that interest signals are an important aspect of driving views, but that the algorithms likely consider other private interest signals that are not available.

**Did it really only take 30 mins to make a viral hit?**

Content creators also rely upon a unique set of internal data social media platforms make available to them that can provide important insights into creating widely disseminated content. Our channel, the Scieneers (Figure 2a), seeks to humanize scientists and make academia accessible to a wide and often underrepresented group of social media users (Figure 1a), most often with lighthearted videos focused on science from an ensemble of researchers at different stages in their academic careers. We also experimented with an "ask us anythings" or "AUA"s, where we answer questions submitted by our audience. While making content, we recorded information on each video such as the time it took to create and also periodically recorded our follower counts, albeit at irregular intervals (Figure 2b). TikTok and YouTube are unique among social media platforms in that they provide some additional data to content creators such as the average time spent by a user watching each video and the percentage of users that watch an entire video.

Categorizing the performance of our 125 videos as low, medium, or high performing in conjunction with using account specific internal data provided us with some key insights about our audience. First, we found that while our follower counts generally increased, not all high performing videos were associated with significant increases in follower counts (Figure 2b). Hence, adding followers does not perfectly correlate with views, and likely also depends on other factors, such as the emotional reaction the video elicited in the audience. We also found that there was no relationship between the videos that performed well versus the amount of time and effort that it took to create the video (Figure 2c). Such data can feel discouraging, as videos that took over a week to produce did not necessarily perform better than videos that took less than thirty minutes to produce. However, this highlights that STEM videos do not necessarily need a significant time investment to reach a large audience. Our analysis also found that the average user will not spend longer than 27 seconds on a video (Figure 2d), and that few videos are ever watched in full, especially as they become longer (Figure 2e). Such data is overall encouraging because it shows that anyone, regardless of time available, can produce high-impact short form STEM-related content.

**Be creative… but also interesting…**

However, there are no hard and fast "rules" with regards to generating and optimizing content for broad dissemination. Our "Disney Princesses as Scientists" series (Videos 1 and 2) took over 1 week for each video to produce and each video was much longer than 27s, and included an ensemble cast (Figure 2f). However, owing to the uniqueness of the content, some of these videos garnered well over 400k views, and were also associated with large increases in our follower counts (Figure 2a). In this case, we believe that our positive message and affirmation of women in STEM strongly resonated with our followers and made them want to see more similar content. As our audience expressed an interest in science related music, we also tried producing a video in which we asked ChatGPT to write a song about science (Figure 2f, Video 3), which we all sung and added music. Unfortunately, despite a week of effort, this video garnered few views. While disappointing, the post-analysis was very informative, as the vast majority of viewers watched the video for <10s, suggesting that our first initial impression wasn't favorable, and highlighting that it is very important to have a "visual hook" for the audience. Our "Ask us anything" series (Figure 2f, Video 4) was an effort to engage with the audience more directly, but consistently had very low views (average <1500). Considering that these videos are largely composed of interview style questions and answers, we were curious as to whether a more interesting visual approach would encourage views and found that overlaying videos of researchers working in the lab in combination with narrated answers to the questions increased our views to >4500. Overall, while we certainly enjoy and will continue to produce some content that requires more effort, we also wanted to highlight that our most popular video (~1M views) took less than 30 mins to create (Figure 2f, Video 5). It's short (9s), has a visual hook (a mystery), and highlighted a humous aspect of academia (the joy of publishing a paper) in a relatable way to the audience (through the use of a popular song).

**Read your fan mail, all of it!**

Computational approaches can also help analyze user comments, a powerful albeit overwhelming aspect of social media. Comments enable the public to directly engage with scientific creators and react to each individual video. While one can certainly scan individual comments, this can be a time-consuming endeavor, especially for large collections of comments. In our case, the Scieneers have received 2050 comments across all our content as of the writing of this work (Figure 3a). To that end, we performed a computational sentiment analysis(*11*) leveraging the Natural Learning Toolkit(*12*) with TextBlob, a python library for processing text, which possesses a lexicon text library that includes different senses of the same word as different entries to unbiasedly classify individual words as positive, neutral or negative (Figure 3b). When calculating the sentiment for a single word, TextBlob takes the average of different entries for the same word. The overall word sentiment across all our videos was 7% negative, 63% neutral, and 30% positive words (Supplementary Figure 3). Using this individual word analysis, we were able to visualize what ideas and concepts were the most relevant to our audience with a word cloud (Figure 3c). This powerful approach allows content creators to determine what themes and concepts most resonate with their audience. For example, words like lab, scientist, women, love, amazing, great, good often appeared in the comments by our audience. The word cloud also enables a targeted approach of key terms to search for if there is a point of concern. As an example, "evil" appeared often, but largely due to a single very engaged follower who was requesting more parody content "Evil scientist POV for Halloween".

Computational approaches to sentiment, although not perfect, can also offer assistance when dealing with colloquial language, which is filled with irony, slang, and sarcasm. As an example, "Got me crying this morning this was so awesome" (Figure 3b), crying is classified as having a negative sentiment, whereas the true sentiment of the comment was positive from unbridled laughter. Hence, it is more important to consider the overall sentiment of the comment. As such we again leveraged TextBlob to classify all whole comments/phrases as positive, neutral, or negative. This is done by taking the average of the individual words in the comment's polarity (positive, negative or neutral). With this methodology, we found that we had a much more positive overall sentiment (Figure 3d). Applying this analysis across categories can also help provide content creators with more detailed information on how their various content types are received by the audience. For example, while our Disney Princess Series or AUAs were not always our most viewed content, the audience that did engage with that content were very enthusiastic about them, as they generated the most positive sentiments of all our comments (Figure 3d).

**Discussion:**

For our fellow academics, our results show that engaging in social media may be more accessible and easier than we ourselves initially perceived. Our group used the creation of social media content as a team building exercise that could also benefit the public, which lowered the stakes for us as we didn't measure our "success" based on views and follower counts. However, in studying our progress, we learned that contributing STEM content to social media doesn't require expensive sets, high end video equipment, or significant investments of time to be successful. Some of our most popular content was created in just 30 minutes, and we often leveraged our scientific environment as a "set". Many social media companies also provide powerful yet simple to use video editing tools, greatly lowering the barriers to starting up.

With regards to each individual video, we also learned that first impressions matter, shorter videos get more views (<27s), unique content is welcome (Disney princess), and maintaining visual interest can be helpful (ask us anything's). It can also be helpful to think about the broader impacts of individual videos. While we all seek to have videos that reach millions of views, many, many videos are still seen hundreds to thousands of times, presumably by different individuals. When put in the perspective of seminar or classroom attendance, there are few rooms on any academic campus that could hold that many people. Hence, every video can matter and make a difference, especially to individuals that don't have access to academic campuses.

Finally, we hope that the computational tools we present here will help streamline gathering feedback on content. These peer-reviewed algorithms are available to all researchers via the open source freely available Python programing language libraries. When looking at your individual videos, we learned that all interest signals matter to one extent or another, although focusing on likes can be very helpful for new content creators. Continuing to contribute and build a following also helps with dissemination, as the most viewed videos in our analysis came from creators with higher follower counts. We encourage you to perform sentiment analysis of comments to better understand what concepts resonate most with the audience and what emotions your content is provoking. Overall, we hope that you will consider joining us in engaging the public with STEM content. Working together, we can help combat misinformation.

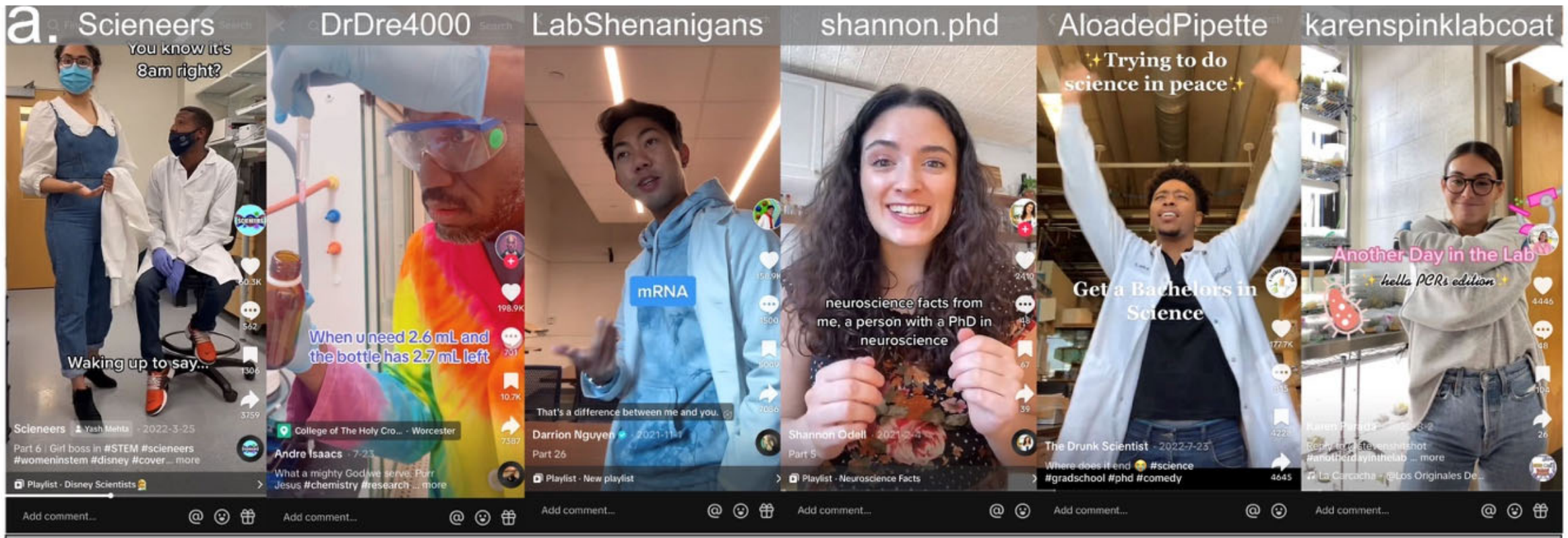

| | | | | | |
|---|---|---|---|---|---|
| **Follower Count:** | | | | | |
| 16.4k | 500.7k | 531.1k | 42.2k | 17.1k | 66.6k |
| **Current Position:** | | | | | |
| Ensemble (Grad Students, Postdoc, Assistant Prof and Research Specialists) | Associate Professor of Chemistry | Self-employed science digital media content creator | Neuroscientist, and Science writer and Editorial Producer | Neuroscientist and Medical writer | Biological Scientists and Clinical Research Coordinator |
| **Thematic Focus:** | | | | | |
| Humanizing scientists and making academia more accessible | Leveraging music and dance to attract the next generation of scientists | Using humor with popular sounds to educate on biological processes | Using humor to educate the masses on mental health and Neuroscience | Demystifying STEAM research with humor | Showcasing the first generation experience in graduate school |

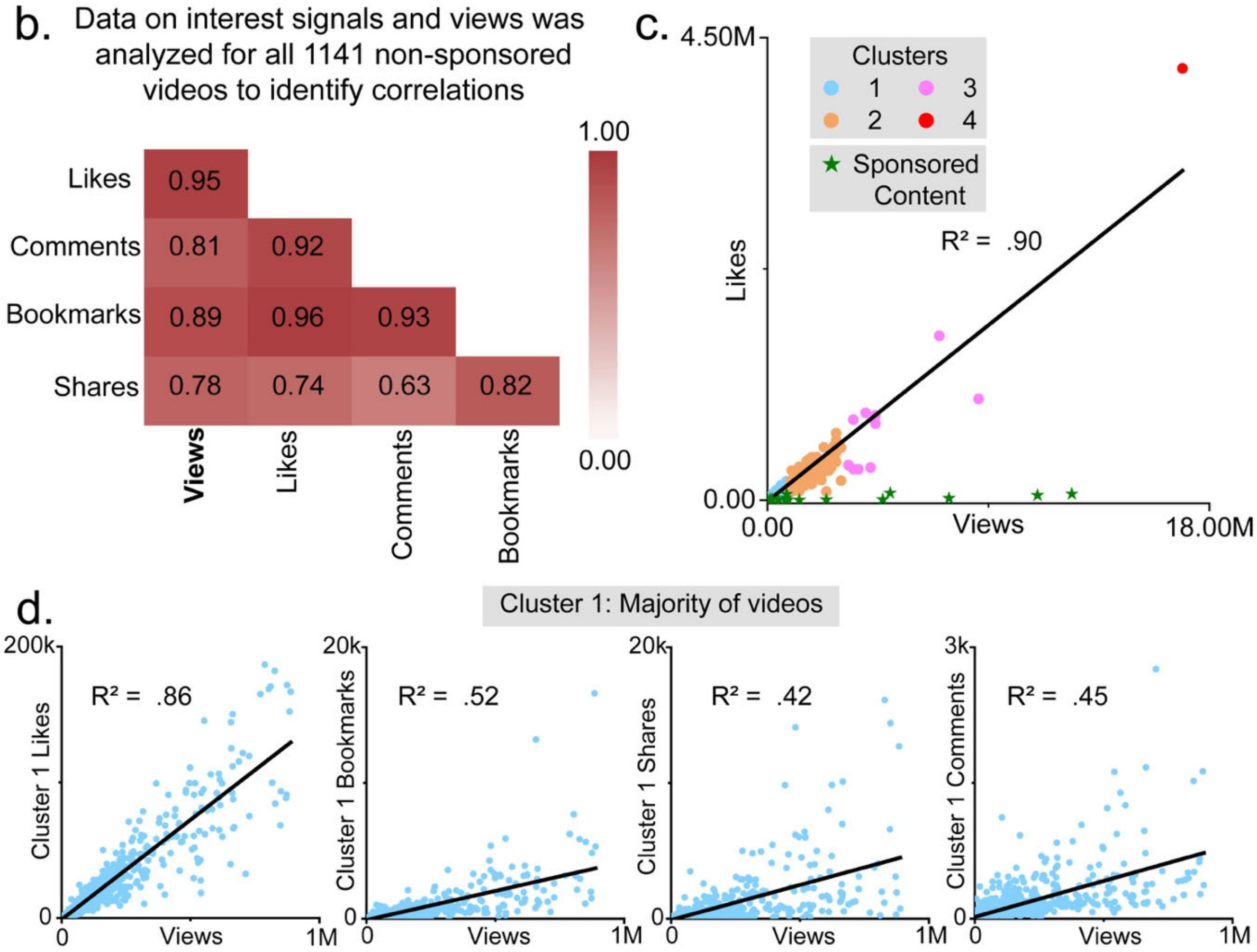

**Figure 1: Correlation, regression and clustering analysis reveals the most rewarded public TikTok interest signal. a.** We analyzed readily accessible TikTok ‘interest signal’ data from 6 scientific creators and 1141 videos with various popularities (Sponsored content not included). **b.** A user can like, comment, bookmark or share videos that they view. The pairwise matrix of Pearson correlation coefficients shows that a viewer liking a video has the strongest correlation to views when compared to other interest signals, although all interest signals appeared to be important. There is also high correlation among interest signals, suggesting that many videos receive more than one signal. **c.** To better understand if all videos are treated the same by the algorithm, clustering analysis was performed, shown here on a linear regression of likes vs views (Sponsored content not included in regression analysis). **d.** The majority of videos fall under cluster 1, especially those from newer creators. In this cluster, likes still strongly correlated with views, but the other interest signals had much weaker correlations than when considering all the videos.

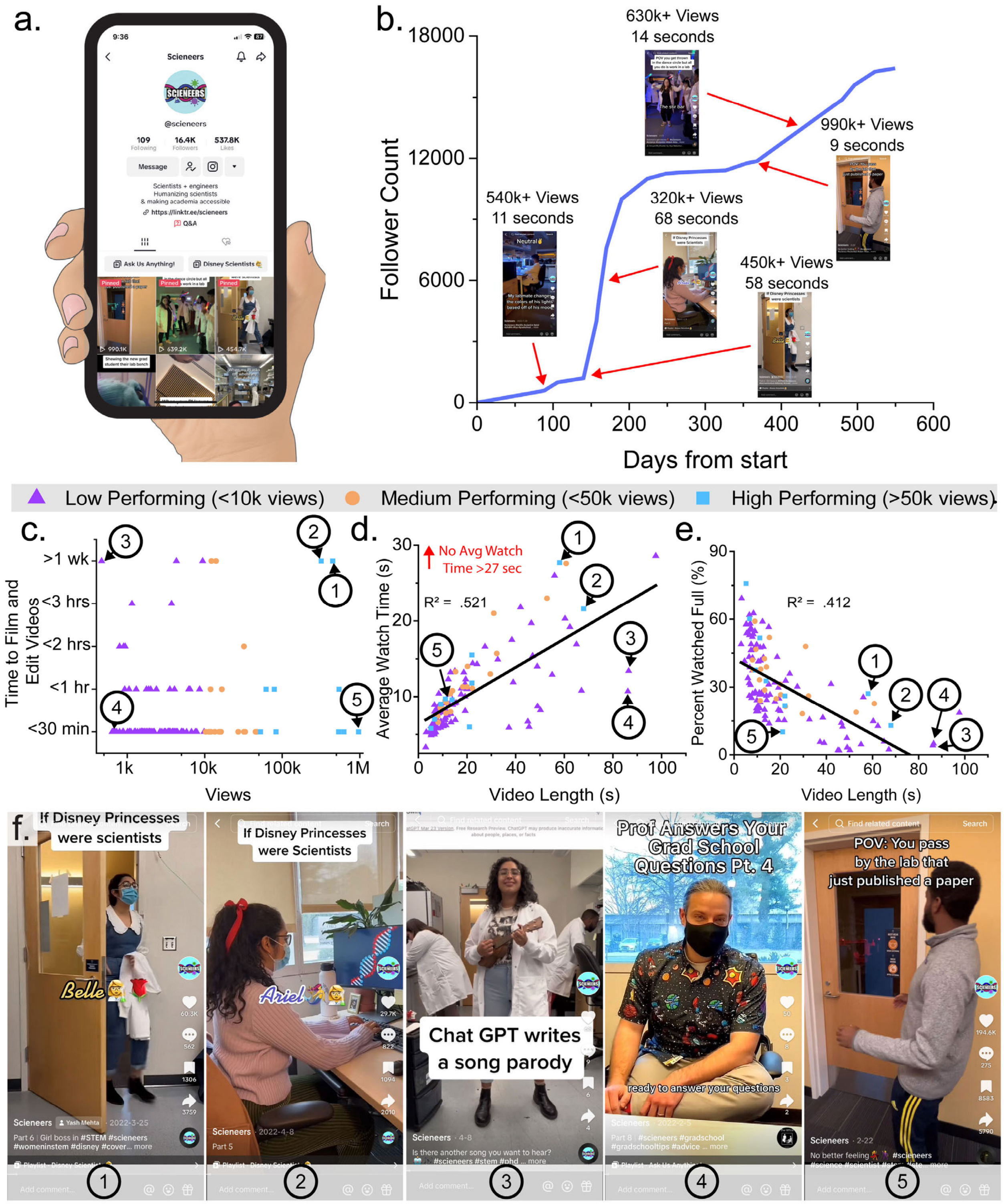

**Figure 2: Analysis of data exclusively available to content creators enables analysis of many video metrics. a.** We created the TikTok account the "Scieneers" in November of 2022 and have amassed over 16,000 followers and 537,000 likes. **b.** Leveraging data provided by TikTok, we were able to track follower growth over time. Interestingly, we observed that videos that garnered the most views did not necessarily equate to the largest follower growth. **c.** In examining the amount of time it took to film a video and edit our videos, we did not find a

correlation with time spent and video performance videos, short and long production times could both lead to highly viewed videos. **d.** We found that regardless of the video length, users do not watch videos (on average) for longer than 27 seconds. **e.** Similarly, the percent of viewers who watch the entire video decreases with video length. **f.** The highlighted 5 videos showcase varying extremes in production time and views. Videos 1 and 2 were a part of our highly popular ‘Disney princess as scientists’ series. They required >1 week of effort, but were very engaging to viewers despite the length. Incidentally, this unique content also led to large increases in our follower counts. Videos 3 and 4 were also longer videos but failed to keep users/viewers engaged, many viewers watched these videos for less than <10s, highlighting the need for a visual hook. Video 5 is an example of a short video, produced in a few minutes, that strongly resonated with users and was highly viewed.

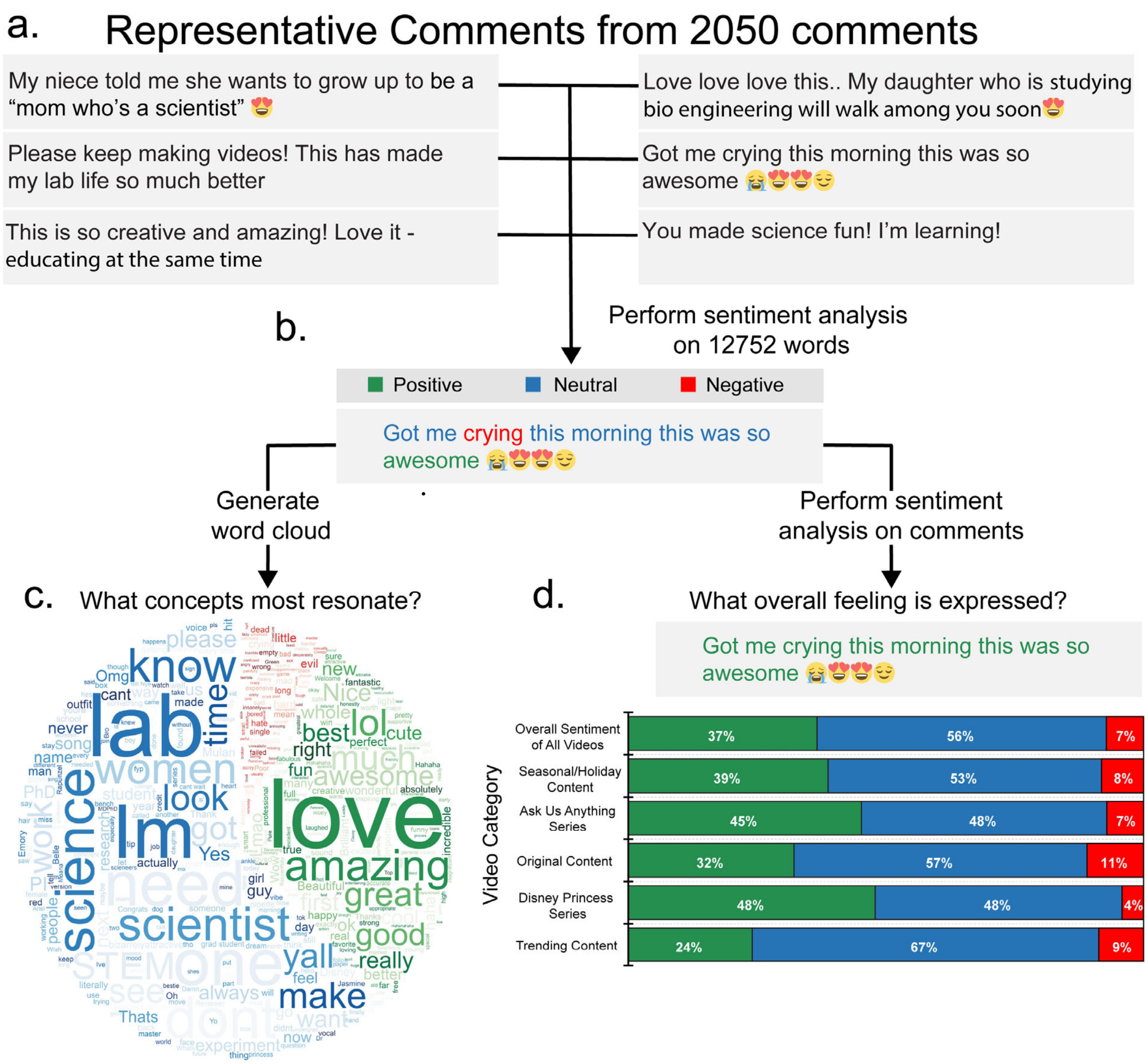

**Figure 3: Sentiment analysis of user comments provides feedback on resonating concepts and evoked feelings. a.** Shown here are select examples of the 2050 individual comments received on the Scieneers account. **b.** We leverage Python's Natural Learning Toolkit (NLTK) with TextBlob to perform a sentiment analysis to analyze user/viewer generated comments to understand what words were most often used by our audience and if these words had a positive, neutral, or negative connotation. **c.** The word cloud shows the most commonly used words in the comments along with the connotation. The larger the word in the word cloud the more it was utilized by our viewers/users. **d.** To understand what overall feeling is expressed in each comment, we again leverage NLTK with TextBlob to analyze entire comments instead of individual words. In general, over 90% of comments were neutral or positive, with our video categories such as 'Disney princess as scientists' and our 'Ask Us Anything series (about grad school)' evoking the most positive emotions with our audience.

**References:**

1. A. K. Isaacs, How to attract the next generation of chemists. *Nature Reviews Chemistry* **7**, 375-376 (2023).
2. D. J. Asai, Race Matters. *Cell* **181**, 754-757 (2020).
3. S. A. Habibi, L. Salim, Static vs. dynamic methods of delivery for science communication: A critical analysis of user engagement with science on social media. *PLoS One* **16**, e0248507 (2021).
4. C. R. Prindle, N. M. Orchanian, L. Venkataraman, C. Nuckolls, Short-Form Videos as an Emerging Social Media Tool for STEM Edutainment. *Journal of Chemical Education* **101**, 1319-1324 (2024).
5. JAMA, in *Most Viewed Articles 2022 – JAMA*. (JAMA, https://jamanetwork.com/journals/jama/pages/2022-most-viewed-jama, 2022), vol. 2024.
6. S. Chew, in *TED Talk,* C. Anderson, Ed. (https://www.ted.com/talks/shou_chew_tiktok_s_ceo_on_its_future_and_what_makes_its_algorithm_different, 2023).
7. W. McKinney, Data Structures for Statistical Computing in Python. (2010).
8. J. Benesty, J. Chen, Y. Huang, I. Cohen, in *Noise Reduction in Speech Processing,* I. Cohen, Y. Huang, J. Chen, J. Benesty, Eds. (Springer Berlin Heidelberg, Berlin, Heidelberg, 2009), pp. 1-4.
9. F. Pedregosa *et al.*, Scikit-learn: Machine Learning in Python. *J. Mach. Learn. Res.* **12**, 2825–2830 (2011).
10. J. A. Hartigan, M. A. Wong, Algorithm AS 136: A K-Means Clustering Algorithm. *Journal of the Royal Statistical Society. Series C (Applied Statistics)* **28**, 100-108 (1979).
11. T. Nasukawa, J. Yi, paper presented at the Proceedings of the 2nd international conference on Knowledge capture, Sanibel Island, FL, USA, 2003.
12. S. Bird, Edward Loper, Ewan Klein, *Natural Languange processing with Python:analyzing text with natural language toolkit.* (O'Reilly Media, Inc, 2009).

**Acknowledgments:** We thank all the users, fans, and creators that help share scientific content on social media platforms. We would like to acknowledge all present and former lab members that were participants in our videos and will be participants in the future.

**Author contributions:** All authors collaborated in the making of the videos and the conceptualization of this work. OO, PD, CAL, AD, MF analyzed videos for other content creators. OO, PD, MF analyzed videos for our account (Scieneers). MF performed the computational analysis for this work. OO made the figures. OO, PD, DRM wrote the manuscript. All authors contributed to editing the manuscript.

**Competing interests:** Authors declare that they have no competing interests.

**Data and materials availability:** Jupyter notebook of methods and codes for computational analysis are available at https://github.com/SL2lab/Oshinowo_et_al_2024

# Supplementary Materials for

# Engineering the communication of science through social media

**Authors:** Oluwamayokun Oshinowo[1,2,3]†, Priscila Delgado[1,2,3]†, Meredith Fay[1,2,3]†, C. Alessandra Luna[1,2,3], Anjana Dissanayaka[1,2,3], Rebecca Jeltuhin[1,2,3], David R. Myers[1,2,3]*

Correspondence to: david.myers@emory.edu,

**This file includes:**

Supplementary Figures 1 to 3

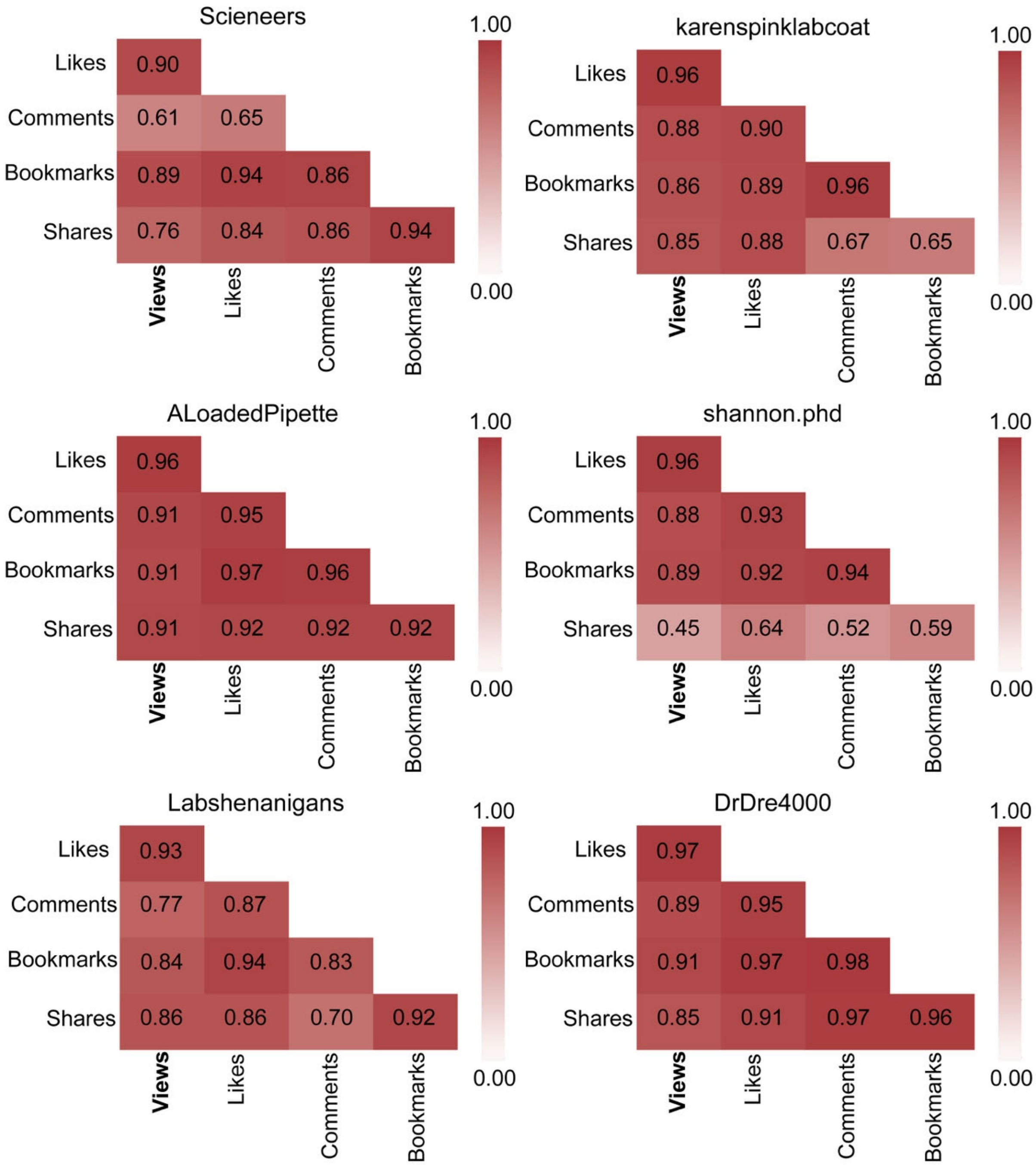

**Supplementary Figure 1**

Correlation matrices highlighting potential differences in interest signal importance between scientific content creators.

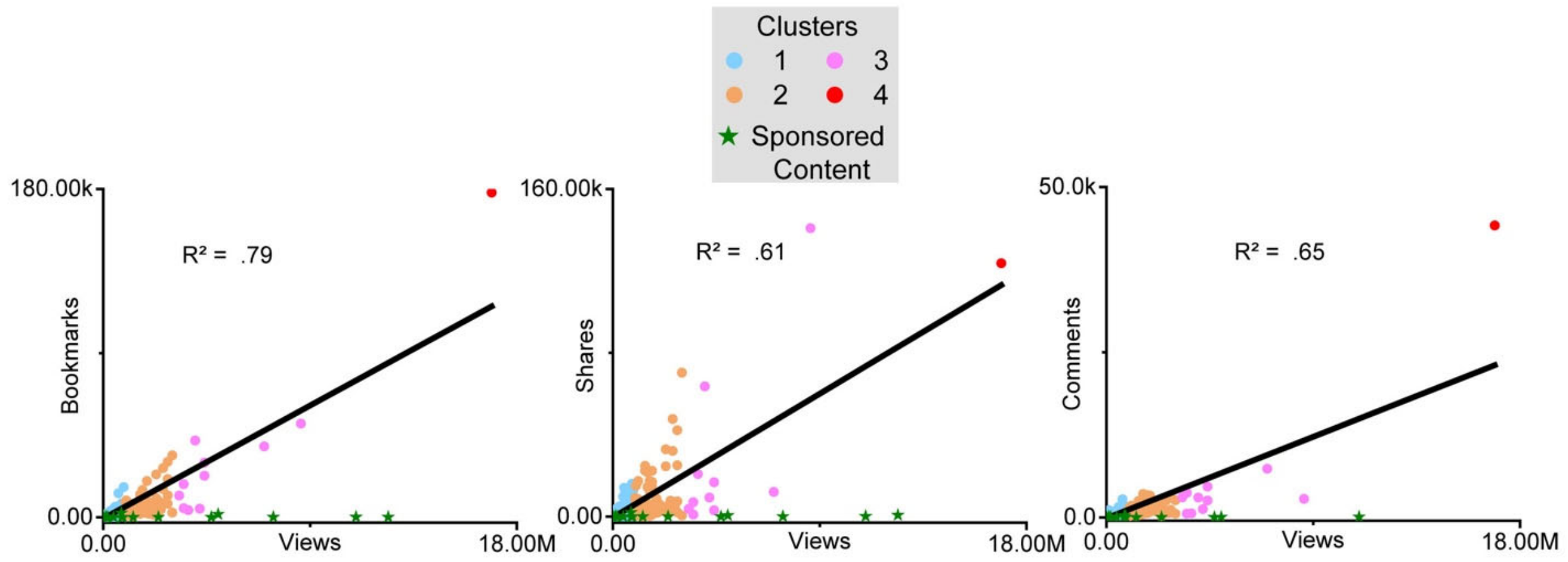

**Supplementary Figure 2**

Cluster and regression analysis of all creator videos with various interest signals (bookmarks, shares and comments to better understand if all videos are treated the same by the algorithm (Sponsored content not included in regression analysis).

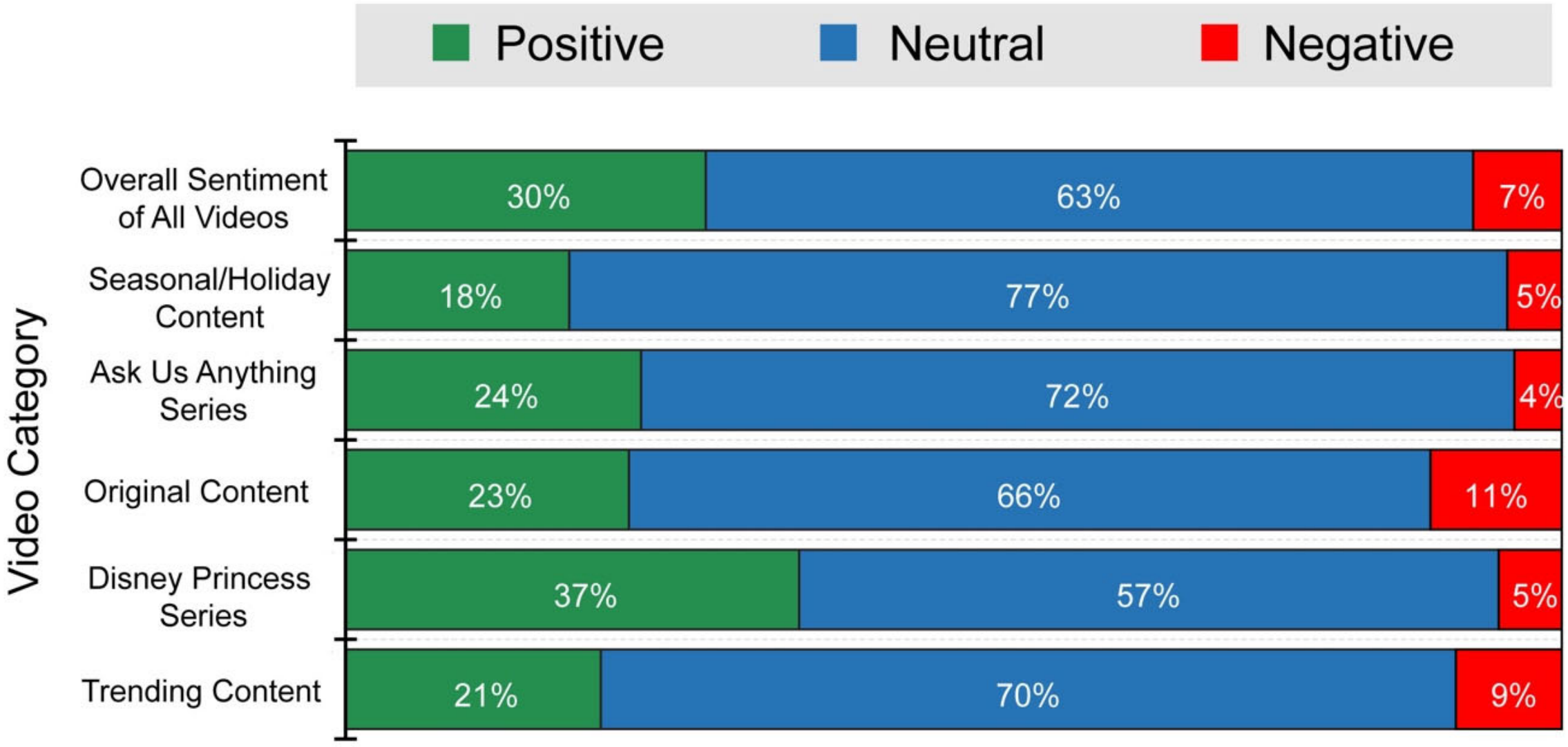

**Supplementary Figure 3**

Sentiment analysis of all individual words in comments.